\documentclass[aps,amssymb,twocolumn,showpacs]{revtex4}
\usepackage{amssymb}
\usepackage{graphicx}
\usepackage{color}
\usepackage{textcomp}
\usepackage{ulem}
\begin{document}

\title{ Chiral crossover characterized by Mott transition at finite temperature }
\author{Shijun Mao}
\affiliation{School of Science, Xi'an Jiaotong University, Xi'an, Shaanxi 710049, China}

\begin{abstract}
We discuss the proper definition for the chiral crossover at finite temperature, based on the Goldstone's theorem. Different from the usually used maximum change of chiral condensate, we propose to define the crossover temperature by the Mott transition of pseudo-Goldstone bosons, which, by definition, guarantees the Goldstone's theorem. We analytically and numerically demonstrate this property in frame of a Pauli-Villars regularized NJL model. In external magnetic field, we find that the Mott transition temperature shows an inverse magnetic catalysis effect.
\end{abstract}

\date{\today}
\maketitle

The change in chiral symmetry is one of the most important properties of quantum chromodynamics (QCD) in hot medium, which is essential to understand the light hadrons at finite temperature and density~\cite{chiralmeson1,chiralmeson2,chiralmeson3,chiralmeson4}. In chiral limit, the phase transition from chiral symmetry breaking in vacuum and at low temperature to its restoration at high temperature happens at a critical temperature $T_c$. In real case with non-vanishing current quark mass, the chiral symmetry restoration is no longer a genuine phase transition but a smooth crossover. Since the crossover does not happen at a point but in a region, the way to describe it with a fixed temperature is not unique. Considering the maximum fluctuations around a continuous phase transition in chiral limit, the pseudo-critical temperature $T_{pc}$ to characterize the chiral crossover in real case is normally defined through the maximum change of the chiral condensate, $\partial^2\langle\bar\psi\psi\rangle/\partial T^2 =0$. From the lattice QCD simulation~\cite{tclqcd}, it is about $T_{pc}\simeq 156$ MeV.

The mechanism for a continuous phase transition is the spontaneous symmetry breaking. One can define an order parameter which changes from nonzero value to zero or vice verse when the phase transition happens. On the other hand, the spontaneous breaking of a global symmetry manifests itself in the Goldstone's theorem~\cite{gold1,gold2}: Whenever a global symmetry is spontaneously broken, massless fields, known as Goldstone bosons, emerge. Corresponding to the spontaneous chiral symmetry breaking, the order parameter is the chiral condensate and the Goldstone modes are pions. If we take $T_{pc}$ defined above as the characteristic temperature of the chiral crossover, the problem is whether the chiral condensate at $T_{pc}$ is already small enough and the pseudo-Goldstone modes at $T_{pc}$ are already heavy enough to guarantee the system to be in chiral restoration phase. According to the Goldstone's theorem, in the chiral breaking phase at low temperature pions as pseudo-Goldstone modes should be in bound states, and in the chiral restoration phase at high temperature pions should be in resonant states with nonzero width. The connection between the two states is the Mott transition at temperature $T_m$~\cite{mott1,mott2,mott3}, where the decay process $\pi\to q\bar q$ starts. It is clear to see $T_{pc}=T_m$ in chiral limit. In real case, however, there is no guarantee for the coincidence of the two temperatures. In this case, the Goldstone's theorem is broken down. Since pions maybe already in resonant states with large mass at $T<T_{pc}$ or still be in bound states with small mass at $T>T_{pc}$. To be consistent with the Goldstone's theorem, we propose to pin down the chiral crossover by the Mott transition of the pseudo-Goldstone boson. Taking into account energy conservation for the decay process, the Mott transition temperature $T_m$ is defined through the pion mass $m_\pi(T)$ and quark mass $m_q(T)$,
\begin{equation}
\label{mott}
m_\pi(T_m)=2m_q(T_m).
\end{equation}

Due to the non-perturbative difficulty in QCD, we calculate the Mott transition temperature in an effective chiral model. One of the models that enables us to see directly how the dynamical mechanism of chiral symmetry breaking and restoration operates is the Nambu--Jona-Lasinio (NJL) model applied to quarks~\cite{njl1,njl2,njl3,njl4,njl5}. Within this model, one can obtain the hadronic mass spectrum and the static properties of mesons remarkably well. In this paper, we calculate in the model the chiral condensate in mean field approximation and the meson mass in random phase approximation (RPA), which is proved to guarantee the Goldstone's theorem in chiral breaking phase.

Recent years, the investigation of chiral symmetry is extended to including external electromagnetic fields. Considering the dimension reduction of fermions in external magnetic fields, the chiral symmetry breaking in vacuum is enhanced by the background magnetic field, from both the lattice QCD simulation~\cite{lattice1,lattice2,lattice3,lattice4,lattice5,lattice6,lattice7} and effective models~\cite{model1,model2,model3,model4,model5}. The surprising is the behavior of the chiral crossover temperature $T_{pc}$. With increasing magnetic field, it decreases from lattice QCD simulations~\cite{lattice1,lattice2,lattice3,lattice4,lattice5,lattice6,lattice7} but increases from effective models in mean field approximation~\cite{model1,model2,model3,model4,model5}. Many scenarios are proposed to understand this qualitative difference between lattice QCD and effective models~\cite{fukushima,kamikado,bf1,bf11,bf13,bf2,bf3,bf4,bf5,bf6,bf7,bf8,bf9,bf10}. We will point out that this difference comes from the definition of the chiral crossover temperature.

The magnatized two-flavor NJL model is defined through the Lagrangian density~\cite{njl1,njl2,njl3,njl4,njl5}
\begin{equation}
\label{njl}
{\cal L} = \bar{\psi}\left(i\gamma_\mu D^\mu-m_0\right)\psi+\frac{G}{2}\left[\left(\bar\psi\psi\right)^2+\left(\bar\psi i\gamma_5\tau\psi\right)^2\right],
\end{equation}
where the covariant derivative $D^\mu=\partial^\mu+iQ A^\mu$ couples quarks with electric charge $Q=diag (Q_u,Q_d)=diag (2e/3,-e/3)$ to the external magnetic field ${\bf B}=B{\bf e}_z$ through the potential $A_\mu=(0,0,Bx,0)$, $G$ is the coupling constant in scalar and pseudo-scalar channels, and $m_0$ is the current quark mass characterizing the explicit chiral symmetry breaking.

Taking the Leung-Ritus-Wang method~\cite{ritus1,ritus2,ritus3,ritus4,ritus5,ritus6}, the chiral condensate $\langle\bar\psi\psi\rangle$ or the dynamical quark mass $m_q=m_0-G\langle\bar\psi\psi\rangle$ at mean field level is controlled by the gap equation
\begin{equation}
\label{gap}
m_q\left(1-GJ_1\right)=m_0
\end{equation}
with
\begin{equation}
J_1 = N_c\sum_{f,n}\alpha_n {|Q_f B|\over 2\pi} \int {d p_z\over 2\pi}{\tanh {E_f\over 2T}\over E_f},
\end{equation}
where $N_c=3$ is the number of colors which is trivial in the NJL model, $\alpha_n=2-\delta_{n0}$ the spin degeneracy, $T$ the temperature of the quark system, and $E_f=\sqrt{p^2_z+2 n |Q_f B|+m_q^2}$ the quark energy with flavor $f=u,d$, longitudinal momentum $p_z$ and Landau energy level $n$.

Mesons in the model are treated as quantum fluctuations above the mean field and constructed through RPA~\cite{njl1,njl2,njl3,njl4,njl5}. In chiral limit with $m_0=0$, the isospin triplet $\pi_0$ and $\pi_\pm$ and isospin singlet $\sigma$ are respectively the Goldstone modes and Higgs mode, corresponding to the spontaneous chiral symmetry breaking at vanishing magnetic field. Turning on the external magnetic field, only the neutral pion $\pi_0$ remains as the Goldstone mode.

With the RPA method, the meson propagator $D_m$ can be expressed in terms of the meson polarization function or quark bubble $\Pi_m$,
\begin{equation}
D_m(q)=\frac{G}{1-G\Pi_m(q)}.
\end{equation}
The meson mass $m_m$ is defined as the pole of the propagator at zero momentum ${\bf q}={\bf 0}$,
\begin{eqnarray}
\label{pole}
1-G\Pi_m(m_m, {\bf 0})=0
\end{eqnarray}
with
\begin{eqnarray}
&& \Pi_m(q_0,{\bf 0}) = J_1-(q_0^2-\epsilon_m^2) J_2(q_0),\\
&& J_2(q_0) = -N_c\sum_{f,n}\alpha_n \frac{|Q_f B|}{2\pi} \int \frac{d p_z}{2\pi}{\tanh {E_f\over 2T}\over E_f (4 E_f^2-q_0^2)},\nonumber
\end{eqnarray}
$\epsilon_{\pi_0}=0$ for the Goldstone mode and $\epsilon_\sigma=2m_q$ for the Higgs mode. In nonzero magnetic field, the three-dimensional quark momentum integration in the gap equation (\ref{gap}) and pole equation (\ref{pole}) becomes a one-dimensional momentum integration plus a summation over the discrete Landau levels.

In chiral limit with vanishing current quark mass, by comparing the gap equation (\ref{gap}) for quark mass with the pole equation (\ref{pole}) for meson mass, we have the analytic solutions
\begin{equation}
m_{\pi_0}=0,\ \ m_\sigma=2m_{q}
\end{equation}
in the chiral breaking phase with $m_q\neq 0$ and
\begin{equation}
m_{\pi_0}=m_\sigma\neq 0
\end{equation}
in the chiral restoration phase with $m_q=0$. A direct consequence of these solutions is that the Mott transition temperature $T_m$ defined by $m_{\pi_0}(T_m)=2m_q(T_m)$ coincides with the critical temperature $T_c$ defined by $m_q(T_c)=0$. The phase transition from chiral symmetry breaking to its restoration is a second order phase transition.

In physical case with nonzero current quark mass, the chiral restoration becomes a smooth crossover. At low temperature, the spontaneous chiral symmetry breaking dominates the system. Considering the fact that the explicit chiral symmetry breaking is slight, we can take $m_0$ expansion in solving the gap equation for quark mass and pole equation for meson mass. With the notation $m_q=m^{cl}_q+\delta_q$ and $m_{\pi_0}=m^{cl}_{\pi_0}+\delta_{\pi_0}$, where $m_q^{cl}$ and $m_{\pi_0}^{cl}=0$ are the quark and neutral pion masses in chiral limit, and keeping only the linear term in $\delta_q$ and quadratic term in $ \delta_{\pi_0}$ in the gap and pole equations, we have
\begin{eqnarray}
\delta_q &=& -\frac{m_0}{m^{cl}_q} \frac{1}{G \frac{\partial J_1}{\partial m_q}{\big |}_{\text cl}},\nonumber\\
\delta_{\pi_0}^2 &=& -\frac{m_0}{m^{cl}_q+\delta_q}\frac{1}{G J_2{\big |}_{{\text cl}}}.
\end{eqnarray}
It is obvious that in chiral limit with $m_0=0$,  we have $\delta_q=0$ and $\delta_{\pi_0}=0$. The explicit chiral symmetry breaking with $m_0 \neq 0$ modifies the dynamical quark mass, and the Goldstone mode in chiral limit becomes a pseudo-Goldstone mode with nonzero mass.

At high temperature, the quark dimension reduction under external magnetic field causes an infrared ($p_z\to 0$) singularity of the quark bubble $\Pi_m(m_m,{\bf 0})$~\cite{maopion1,maopion2,pion3}. For the pseudo-Goldstone mode $\pi_0$, the infrared singularity of $\Pi_{\pi_0}(m_{\pi_0},{\bf 0})$ happens at the Mott transition temperature $T_m$ where the mass $m_{\pi_0}$ jumps up from $m_{\pi_0}<2m_q$ to $m_{\pi_0}>2m_q$. This indicates a sudden transition from a bound state to a resonant state~\cite{maopion1,pion3}.

Now we do numerical calculations on the Mott transition temperature in both chiral limit and real case. Because of the four-fermion interaction, the NJL model is not a renormalizable theory and needs a regularization. To guarantee the law of causality in magnetic fields, we apply the Pauli-Villars regularization scheme as explained in detail in Ref.~\cite{bf10}. The three parameters in the NJL model, namely the current quark mass $m_0$, the coupling constant $G$, and Pauli-Villars mass parameter $\Lambda$ are listed in Table \ref{table1} by fitting the chiral condensate $\langle\bar\psi\psi\rangle$, pion mass $m_\pi$ and pion decay constant $f_\pi$ in vacuum at $T=B=0$. We take current quark mass $m_0=0$ in chiral limit and $6.4$ MeV in real world.
\begin{table}[htbp]
\centering
\caption{The NJL parameters in Pauli-Villars regularization.}
\label{table1}
\begin{tabular}{cccccc}
\hline
\hline
$m_0$ & $G$ & $\Lambda$ & $\langle\bar\psi\psi\rangle$ & $m_\pi$ & $f_\pi$ \\
(MeV) & (GeV)$^{-2}$ & (MeV) & (MeV)$^3$ & (MeV) & (MeV) \\
\hline
0 &5.03 &977.3 &-230$^3$  &0       &93\\
\hline
6.4 &4.9 &977.3 &-230$^3$  &134   &93\\
\hline
\hline
\end{tabular}
\end{table}

\begin{figure}[hbt]
\centering
\includegraphics[width=7cm]{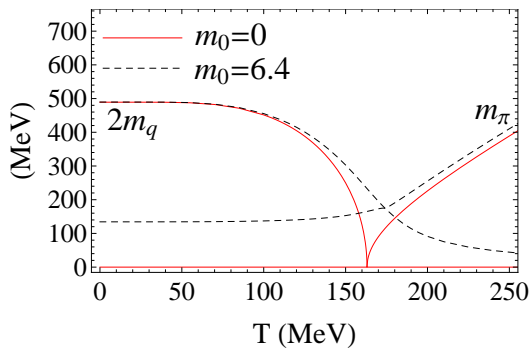}
\caption{The dynamical quark mass $m_q$ and pion mass $m_\pi$ as functions of temperature in chiral limit (solid lines) with current quark mass $m_0=0$ and real world (dashed lines) with $m_0=6.4$ MeV. }
\label{fig1}
\end{figure}

Let's firstly discuss the temperature behavior of the phase structure for chiral symmetry at $B=0$. Fig.\ref{fig1} shows the quark mass and pion mass as functions of temperature in chiral limit and real world. In vanishing magnetic field, the three pions are all Goldstone or pseudo-Goldstone modes. To clearly see the Mott transition temperature $T_m$ and its difference from the pseudo-critical temperature $T_{pc}$, we plot $2m_q$ instead of $m_q$ itself. In chiral limit, the quark mass which is proportional to the order parameter $\langle\bar\psi\psi\rangle$ continuously decreases at low temperature, reaches zero at the critical temperature $T_c=163$ MeV, and keeps zero at higher temperature. This means a second order chiral phase transition. Correspondingly, the Goldstone modes $\pi$ keep massless in the chiral breaking phase and start to be massive at $T_c$. It is clear to see that the Mott transition temperature defined through the threshold condition (\ref{mott}) is exactly the critical temperature $T_m=T_c=163$ MeV. In real world, the quark mass continuously drops down and the pion mass continuously goes up in the whole temperature region. In this case, the chiral phase transition becomes a smooth crossover, and there is no strict definition for the crossover temperature. Generalizing the idea of maximum fluctuations around the second order phase transition in chiral limit, people usually use the maximum change of the chiral condensate or dynamical quark mass to identify the crossover, and refer the corresponding temperature as the pseudo-critical temperature $T_{pc}$ of the chiral crossover. From the definition $\partial^2m_q/\partial T_{pc}^2=0$, we have numerically $T_{pc}=162$ MeV which is close to the lattice QCD result ($156$ MeV~\cite{tclqcd}). From the definition $m_\pi(T_m)=2m_q(T_m)$ and denoted by the crossing point of the two dashed lines in Fig.\ref{fig1}, the Mott transition temperature is different from the pseudo-critical temperature, $T_m=174$ MeV $>T_{pc}=162$ MeV. This means that, pions as pseudo-Goldstone modes can still survive as bound state after the chiral crossover characterized by $T_{pc}$. This explicitly breaks down the Goldstone's theorem.

\begin{figure}[hbt]
\centering
\includegraphics[width=7cm]{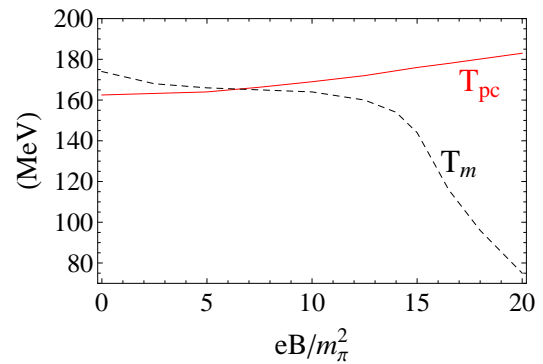}
\caption{ The pseudo-critical temperature $T_{pc}$ (solid line) and Mott transition temperature $T_m$ (dashed line) for the pseudo-Goldstone mode as functions of magnetic field in physical case with $m_0=6.4$ MeV. }
\label{fig2}
\end{figure}

When the external magnetic field is turned on, from lattice QCD simulations~\cite{lattice1,lattice2,lattice3,lattice4,lattice5,lattice6,lattice7} the pseudo-critical temperature $T_{pc}$ drops down with increasing magnetic field, called inverse magnetic catalysis. This confused the people who obtained magnetic catalysis effect from many effective models~\cite{model1,model2,model3,model4,model5}, including the NJL model in mean field approximation. This confusion will disappear when we take $T_m$ instead of $T_{pc}$ to characterize the chiral crossover. Since charged pions interact with the magnetic field, they are no longer pseudo-Goldstone modes, and only neutral pion is the pseudo-Goldstone boson corresponding to spontaneous chiral symmetry breaking. Fig.\ref{fig2} shows the temperatures $T_{pc}$ and $T_m$ as functions of magnetic field. While $T_{pc}$ is controlled by the magnetic catalysis, namely it increases with increasing magnetic field, the Mott transition temperature $T_m$ shows clearly the inverse magnetic catalysis effect in the whole magnetic field region. The physics to explain this difference is the following: $T_{pc}$ is controlled by quarks which are normally calculated at mean field level, while $T_m$ is governed by both quarks and mesons, and the latter are treated as quantum fluctuation above mean field. It is the quantum fluctuation which changes magnetic catalysis to inverse magnetic catalysis.

Again the result of $T_{pc}\neq T_m$ breaks down the Goldstone's theorem in any magnetic field. In weak magnetic field with $eB/m_\pi^2<7$, where $m_\pi$ is the pion mass in vacuum at $T=B=0$, there is $T_m>T_{pc}$ which leads to still surviving neutral pion as bound state in chiral restoration phase. On the other hand, in strong magnetic field with $eB/m_\pi^2 > 7$, there is $T_m < T_{pc}$ which results in already disappeared neutral pion in chiral breaking phase. The degree of the Goldstone's theorem breaking is significantly large in strong magnetic field.
\begin{table}[htbp]
\centering
\caption{Modeling the lattice simulation in NJL calculation.}
\label{table2}
\begin{tabular}{cccc}
\hline
\hline
$eB/m^2_\pi$  \  & $G(B)/G(0)$   \  & $T_{pc}(B)/T_{pc}(0)$   \  & $T_{m}(B)/T_m(0)$  \\
\hline
0     \         &  1      \   & 1    \      &  1     \\
\hline
10     \        &  0.97  \    & 0.99    \      & 0.88     \\
\hline
20     \        &  0.9   \   & 0.97     \     & 0.39      \\
\hline
\hline
\end{tabular}
\end{table}

The above results on $T_{pc}$ and $T_m$ are from the NJL model. A natural question then arises: Whether the breaking down of the Goldstone's theorem comes from the magnetic catalysis in the model which is inconsistent with the lattice QCD simulation. Since the pion mass is not yet available, it is now difficult to answer this question in the frame of lattice simulation. However, we can model the lattice result by embedding the inverse magnetic catalysis in the NJL calculation. The inverse magnetic catalysis, namely a decreasing $T_{pc}$ in magnetic field, should be resulted from a weakening interaction among quarks. We introduce a magnetic field dependent coupling constant $G(B)$ in the gap equation (\ref{gap}) and fix it by fitting the lattice simulated $T_{pc}(B)/T_{pc}(B=0)$~\cite{lattice1}. With this $B$-dependent coupling, we recalculate the quark mass $m_q$, pion mass $m_{\pi}$, and the Mott transition temperature $T_m$. The result of this modeling is shown in Table \ref{table2} at magnetic field $eB/m_\pi^2=0, 10$ and $20$, where $T_{pc}(B)/T_{pc}(B=0)$ is the input from the lattice QCD simulation, $G(B)$ is the output of the gap equation (\ref{gap}), and $T_m$ is the output of the pole equation (\ref{pole}). Different from Fig.\ref{fig2} where $T_{pc}$ increases but $T_m$ decreases with magnetic field, both $T_{pc}$ and $T_m$ in Table.\ref{table2} decrease with magnetic field. Therefore, independent of magnetic catalysis or inverse magnetic catalysis for the pseudo-critical temperature $T_{pc}$, the Mott transition temperature $T_m$ is different from $T_{pc}$ and shows inverse magnetic catalysis effect.

We investigate in this Letter the chiral crossover at finite temperature and in external magnetic fields. In physical world, the chiral restoration is a smooth crossover due to the explicit chiral symmetry breaking. Different from the usually used maximum change of chiral condensate, we propose to define the crossover temperature by the Mott transition of pseudo-Goldstone bosons. This, by definition, guarantees the Goldstone's theorem for the chiral symmetry. As an analytical example, we calculate the order parameter (dynamical quark mass) and Goldstone mode (pion mass) in frame of a Pauli-Villars regularized NJL model. If we take the maximum change of chiral condensate to describe the chiral crossover, the pseudo-Goldstone mode will still survive in the chiral restoration phase in weak magnetic fields and already disappear in the chiral breaking phase in strong magnetic fields. This breaking down of the Goldstone's theorem is independent of the magnetic catalysis or inverse magnetic catalysis for $T_{pc}$. To precisely fix the Mott transition temperature as the characteristic temperature of the chiral crossover, which shows inverse magnetic catalysis effect, we need the result from lattice QCD simulations.\\

\noindent {\bf Acknowledgement:}
The work is supported by the NSFC Grant 11775165 and Fundamental Research Funds for
the Central Universities. We thank the hospitality of Prof. Dirk H. Rischke in Frankfurt University. Part of the work is done when I am visiting his group as the EMMI visiting professor. We also acknowledge partial support by the
``Extreme Matter Institute'' EMMI funded by the Helmholtz Association.

\end{document}